\documentclass[twocolumn,showpacs,preprintnumbers,amsmath,amssymb]{revtex4}
\usepackage{dcolumn}
\usepackage{bm}
\usepackage[dvipdf]{graphicx}
\DeclareGraphicsExtensions{.jpg,.pdf,.mps,.png,.eps,.ps,.EPS}
			   \begin{document}
\def\be{\begin{equation}}
\def\ee{\end{equation}}
\def\bc{\begin{center}}
\def\ec{\end{center}}
\def\bea{\begin{eqnarray}}
\def\eea{\end{eqnarray}}

\title{ Number of $h$-cycles in the Internet at the Autonomous System Level}
\author{Ginestra Bianconi$^1$, Guido Caldarelli$^{2,3}$ and Andrea Capocci$^{3,4}$}
\affiliation{$^1$The Abdus Salam International Center for Theoretical Physics, Strada Costiera 14, 34014 Trieste, Italy \\
$^2$INFM UdR Roma1 and Dipartimento di Fisica Universit\'a ``La Sapienza'',
P.le A. Moro 2, 00185 Roma, Italy\\
$^3$Centro Studi e Ricerche E. Fermi, Compendio Viminale, Roma, Italy\\
$^4$Section de Physique, Universit\'e de Fribourg, P\'erolles 1700 Fribourg, Switzerland}

\begin{abstract}
We present here a study of the clustering and cycles present in the 
graph of Internet at the Autonomous Systems level.
Even if the whole structure is changing with time, we present some evidence that 
the statistical distributions of cycles of order 3,4,5 remain stable during the 
evolution. This could suggest that cycles are among the 
characteristic motifs of the Internet. 
Furthermore, we compare data with the results obtained for growing network models 
aimed to reproduce the Internet evolution. 
Namely the fitness model, the Generalized Network Growth model and the Bosonic Network model.
We are able to find some qualitative agreement with the experimental situation 
even if the actual  number of cycles seems to be larger in the data than in any 
proposed growing network model. The task to capture this feature of the 
Internet represent one of the  challenges in the future Internet modeling.
\end{abstract}
\maketitle
Internet is a beautiful example of a complex system with many 
degrees of freedom resulting in global scaling properties.  
It has been shown \cite{Faloutsos,calda} that the Internet belongs to 
the wide class of scale-free networks \cite{RMP,Strogatz,DoroRev,amaral00}. 
Indeed, it can be described as a network, with nodes and links
representing respectively Autonomous Systems (AS) and physical lines
connecting them; moreover, its degree distribution follows a
power-law behavior.

Different topological quantities have also been measured beside the
degree distribution exponent.
Among those, the clustering coefficient $C(k)$ and the average nearest
neighbor degree $k_{nn}(k)$ of a node as a function of its degree $k$
\cite{Vespignani1,Vespignani2,krapivsky01}. 
In particular, measurements in Internet yield $C(k)\sim k^{-0.75}$
\cite{Vespignani_ps} and $k_{nn}~\sim k^{-\nu}$ with
$\nu\simeq0.5$ \cite{Vespignani_ps}. A two-vertices degree
anti-correlation has also been measured \cite{Maslov}.  
Accordingly, Internet is said to display disassortitative
mixing \cite{Newman_mixing}, because nodes prefer to be linked to peers
with different rather than similar degree. 
Moreover, the modularity of the Internet due to the 
national patterns has been studied by measuring the slow
decaying modes of a diffusion process defined on it \cite{Maslov_dm}. 

Recently, more attention has been devoted to network motifs
\cite{UAlon,Milo}, i.e. subgraphs that recur with a
higher frequency than in maximally random graphs with the same degree
distribution.
Among those, the most natural class includes
cycles\cite{Loops,Guido_cycles}, closed paths of various lengths that
visit each node only once.
Cycles (or loops) are interesting because they account for the 
multiplicity of paths between any two nodes. Therefore, they encode  
the redundant information in the network structure. 
Following the arguments of \cite{UAlon}, it can be shown that the number $N_h$ of 
cycle of size $h$, in a equilibrium undirected scale-free network of 
$N$ nodes with a power-law degree distribution $P(k)\sim k^{-\gamma}$, is
\be \label{powlaw}
N_h(N)\sim N^{\xi(h)}
\ee
with 
\begin{equation}
 \xi(h) = \left\{ \begin{matrix} 
1 & \mbox{ for} &\gamma\leq 2 \cr 
3-\gamma & {\mbox{ for }} &2<\gamma\leq 3 \cr
0 & \mbox{ for} &\gamma\geq 3 \end{matrix}.
\right.
\label{xi.eq}
\end{equation}
In other words, $N_h(N)$ is an algebraic function of the system size with an 
exponent $\xi$ independent of the length $h$ of the cycle.

In contrast, the only analytical result \cite{Loops} for off-equilibrium,
scale-free networks refers to the Barab\'asi-Albert 
model \cite{BA}, and reads
\be \label{pow-log}
N_h(N)\sim\left( \frac{m}{2} log(N)\right)^{\psi(h)},
\ee
with $\psi(h)=h$.

To measure the actual scaling in Internet at the AS level, 
we considered its symmetrical adjacency matrix $\{a_{ij}\}$, with
$a_{ij} =1$ if $i$ and $j$ are connected and $a_{ij}=0$ otherwise. 
We assume that no self-loop is present, i.e.
$a_{ii}=0$ for all $i$. 
In this case, for $h=3$ we simply have \cite{Loops}
\be
N_3=\frac{1}{6}\sum_i (a^3)_{ii}.
\label{n3}
\ee 
For  $h=4$ and $h=5$, by simple arguments it is possible to show that
\be
N_4=\frac{1}{8}\left[\sum_i (a^4)_{ii}-2\sum_i(a^2)_{ii} (a^2)_{ii}+\sum_i (a^2)_{ii}\right]
\label{n4}
\ee
and that
\be
N_5=\frac{1}{10}\left[\sum_i (a^5)_{ii}-5\sum_i(a^2)_{ii} (a^3)_{ii}+5\sum_i (a^3)_{ii}\right].
\label{n5}
\ee

The data of the Internet at the Autonomous System level are collected by the 
University of Oregon Route Views Project and made available by the 
NLANR (National Laboratory of Applied Network Research). 
The subset we used in this manuscript 
are mirrored at  COSIN web page http://www.cosin.org.
We considered 13 snapshots of the Internet network at the AS level at
different times starting from November 1997 (when $N=3015$) toward
January 2001 ($N=9048$). 
Throughout this period, the degree distribution is a power-law
with a nearly constant exponent $\gamma \simeq 2.22(1)$.
Using relations $(\ref{n3})$, $(\ref{n4})$, $(\ref{n5})$, we measure
$N_h(t)$ for $h=3,4,5$ in the Internet at different times,
corresponding to different network size.
\begin{figure}\includegraphics[width = 75 mm, height = 75 mm ]{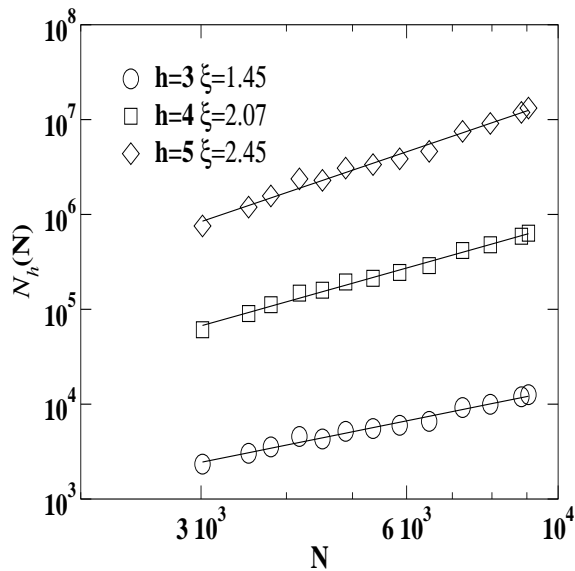} 
\caption{Number of $h$-loops $N_h$ as a function of the system size $N$ 
for loops of length 3,4,5.} 
\label{AS}
\end{figure}
We observe in figure \ref{AS} that the data follow a scaling of the type (\ref{powlaw}),
as predicted by \cite{UAlon} for maximally random (equilibrium)
scale-free networks.
Unfortunately, the exponents $\xi(h)$ strongly depend on $h$, as
reports table \ref{xi_table}, and significantly exceed the predicted
value (Eq.$(\ref{xi.eq})$) for equilibrium scale-free networks with
same $\gamma$, that is, $\xi=0.78$.

So, we can state that loops up to size $5$ are much more frequent in
Internet than in a random scale-free networks with a similar degree
distribution.  
$\xi(h)$ and $N_h$ are large even when compared with off-equilibrium
networks inspired by the Internet growth. 
The models we consider here reproduce the most accurately the Internet 
behavior as regards the degree, clustering and centrality 
probability distributions.

The fitness model \cite{Fitness}, for example, is a growing network
model where, at each time step, a new node is added to the network
and connected by $m$ links to existing ones.
Each node has a fitness $\eta_i$, randomly drawn from a uniform distribution 
in $[0,1]$, which enters into the probability that a node acquires a new link, 
\be
\Pi_i=\frac{\eta_i k_i(t)}{\sum_j \eta_j k_j(t)}.
\ee
The fitness represents an intrinsic ability of a node in the
acquisition of new links
The resulting network is a scale-free one with $\gamma=2.255$. 
It has also been found \cite{Vespignani1,Vespignani2} that $C(k)$ and
$k_{nn}(k)$ are in qualitative agreement with Internet data.

As a second instance, we compare the Internet data to the recently
proposed Generalized Network Growth Model (GNG)
\cite{Internet_model}. According to the its definition, at each time step 
\begin{enumerate}
\item{either a node is added and linked with vertex $i$ with probability}
\begin{equation}
p\frac{k_i}{\sum_{j=1,N} k_j}.
\end{equation}
\item{or a link is added (if absent) between nodes $i$ and $j$
already present. with probability}
\begin{equation}
(1-p)\frac{k_i}{\sum_{k=1,N} k_k}\frac{|k_i-k_j|}{\sum_{k\neq i=1,N} |k_i-k_k|}.\end{equation}
\end{enumerate}
The resulting network is a scale-free one, with $\gamma(p)=2+\frac{p}{2-p}$.
Besides, it displays the non trivial features of the degree
correlations as measured in Internet.

Finally, we considered the Bosonic Network (BN), where each node $i$ is assigned an 
innate quality in the spirit of Ref.\cite{capcal}, represented by a
random 'energy' $\epsilon_i$ drawn from the probability distribution
$p(\epsilon_i)$. 
The attractiveness of each node $i$ is then determined jointly by its
connectivity $k_i$ and its energy $\epsilon_i$. Namely, the
probability that node $i$ acquires a link at time $t$ is given by 
\be
\Pi_i=\frac{e^{-\beta\epsilon_i}k_i(t)}{\sum_je^{-\beta\epsilon_j}k_j(t)},
\ee 
i.e. low energy, high degree nodes are more likely to acquire new links.
The parameter $\beta=1/T$ in $ \Pi_i$ tunes the relevance of the quality with respect 
to the degree in the acquisition probability of new links.
Indeed, for $T\rightarrow \infty$ the probability $\Pi_i$ does not
depend any more on the energy $\epsilon_i$ and the BN model reduces to
the Barab\'asi-Albert (BA) model, based only on preferential
attachment.

On the other hand, in the limit $T \rightarrow 0$ only the lowest
energy node has non-zero probability to acquire new links. 
In Ref. \cite{bose}, it has been shown that the connectivity
distribution in this network model can be mapped into the occupation
numbers of a Bose gas. Accordingly, one would expect a corresponding
phase transition for the topology of the network at some temperature
value $T_c$.
In fact, such a critical value is observed for energy distributions where 
($p(\epsilon)\rightarrow 0$ for $\epsilon \rightarrow 0$).
For $T>T_c$ the system is in the ``fit-get-rich''(FGR) phase, 
where low-energy nodes acquire links at a higher rate that
high-energy ones, while for $T<T_c$ a ``Bose-Einstein
condensate''(BEC) or ``winner-takes-all'' phase emerges, where a
single nodes grabs a finite fraction of all the links.
We simulated this model assuming   
\be
p(\epsilon)=(\theta+1)\epsilon^\theta\ \   \mbox{and} \ \  \epsilon\in (0,1) 
 \label{p_epsilon}
\ee
where $\theta=0.5$. 
Varying $T$, one observes a change in the behavior of $N_h$ in the bosonic network
from a scaling of the type (\ref{pow-log}), shown to be exact in the
$\beta=0$ limit for the BA network model \cite{Loops}, 
to a scaling of the type (\ref{powlaw}), valid in the low-temperature limit.
In reference \cite{Loops}, we claim that this change occurs right at
the Bose-Einstein condensation temperature $T_c$.
A careful analysis of the transition shows in fact that the transition is 
rather smooth at $T_c$.

In order to compare networks with a similar mean degree ($<k>=3.5$ for
the Internet), we consider the fitness model with $m=2$ ($<k>=2m=4$)
and the GNG model with parameter $p=0.5$ ($<k>=2/p=4$) and $p=0.6$ 
($<k>=2/p=3.33$). In the GNG network with $p=0.5,0.6$
one numerically finds $\gamma=2.5(2)$ \cite{Internet_model}.
\begin{figure}
\includegraphics[width = 88 mm, height = 40 mm ]{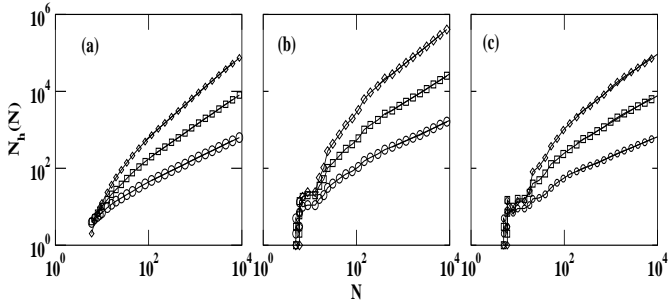} 
\caption{Number of $h$-loops $N_h$ with $h=3,4,5$ in  fitness model with 
$m=2$ (graph(a)) and GNG model with $p=0.5$ (graph(b)) and $p=0.6$ (graph(c)) of 
size up to $N=10^4$. The data asymptotically follow the scaling 
($\ref{powlaw}$) with exponents that remain well below the Internet data.} 
\label{fitness.fig}
\end{figure}

In figure $\ref{fitness.fig}$, we show the scaling of $N_h$ as a function of 
the system size for the fitness model with $m=2$ and the GNG model with 
$p=0.5$, $p=0.6$. For large $N$, $N_h(N)$ is a power-law as in the real 
Internet, yet with much smaller exponents, as shown in Table \ref{xi_table}.

\begin{table}[h] 
\begin{center}
\begin{tabular}{|c|c|c|c|}
\hline
System & $\xi(3)$ & $\xi(4)$ & $\xi(5)$ \\
\hline
AS & $1.45 \pm 0.07$ & $2.07 \pm 0.01$ & $2.45 \pm 0.01$ \\
Fitness & $0.59 \pm 0.02$ & $0.86 \pm 0.02$ & $1.10 \pm 0.02$ \\
GNG (p=0.5) & $0.53 \pm 0.03$ & $0.72 \pm 0.03$ & $0.96 \pm 0.02$ \\
GNG (p=0.6) & $0.53 \pm 0.03$ & $0.74 \pm 0.03$ & $0.99 \pm 0.02$ \\
\hline
\end{tabular}
\end{center}
\caption{The exponent $\xi(n)$ for $n=3,4,5$ as defined in
equation (\ref{powlaw}) for real data and network models.}
\label{xi_table}
\end{table}

When considering the bosonic network model, the picture is more complicated. 
The loops number behavior depends strongly on the temperature parameter.

We can distinguish a high-temperature phase, where $N_h(N)$ is
better fitted by $(\ref{pow-log})$- FGR phase- and a low-temperature
phase, where $N_h(N)$ 
scales as $(\ref{powlaw})$ - BEC phase.
\begin{figure}
\includegraphics[width = 85 mm, height = 45 mm ]{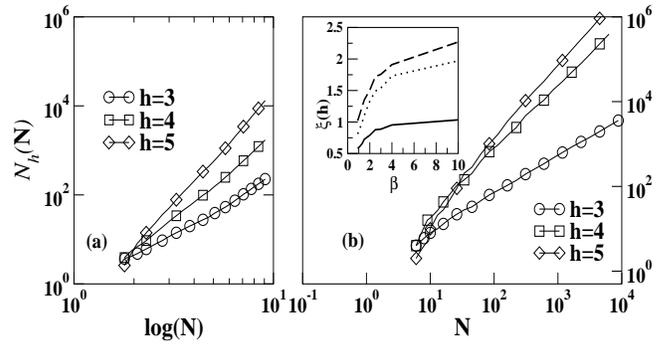}
\caption{The number of cycles in a bosonic network, for (a) $\beta=0.5$
  and (b) $\beta=2.5$. In the inset of (b) we plot the exponents $\xi(h)$, 
for $h=3$ (solid line),$4$ (dotted line),$5$ (dashed line) 
as a function of the inverse temperature $\beta$.} 
\label{exp.fig}
\end{figure}
Even when one decreases the temperature, $\xi(h)$ remains always
far from the real network exponents, as it is shown in figure
\ref{exp.fig}, so that also the bosonic
network fails in reproducing correctly such feature.
Furthermore, no significant sign for a 'winner' node are found in the 
Internet data in which the most connected node has a fraction of links 
$k/N=2024/9048=0.22$ for the January 2001 AS data.

Following \cite{Guido_cycles}, we also measured the clustering coefficients $c_{3,i}$ and
$c_{4,i}$ as a function of the connectivity $k_i$ of node $i$ for all $i$'s.
In particular, $c_{3,i}$ is the usual clustering coefficient $C$, i.e. the number of 
triangles including node $i$ divided by the number of possible triangles $k_i(k_i-1)/2$.

\begin{figure}
\includegraphics[width = 88 mm, height = 88 mm ]{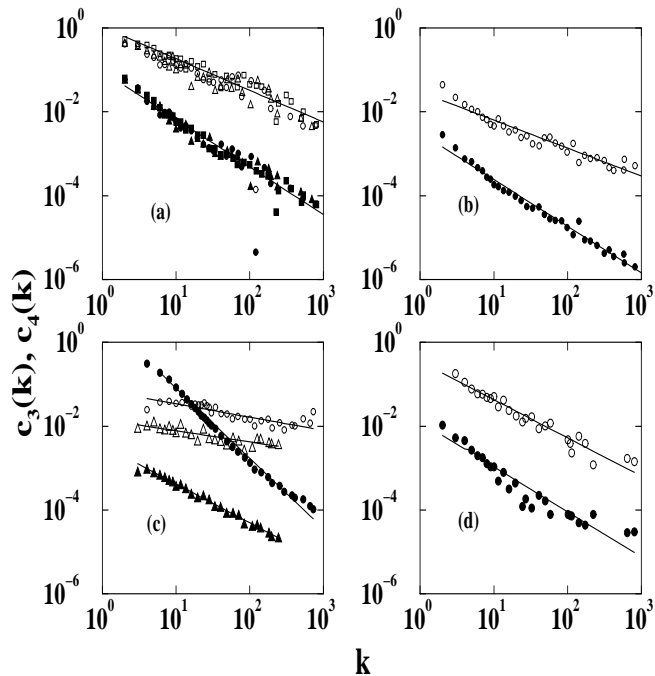} 
\caption{The clustering coefficients $c_3(k)$ and $c_4(k)$ in Internet 
(graph(a)) and in the fitness model (graph(b)), 
the GNG model (graph(c)) with $p=0.5$ (circles), $p=0.6$( triangles)
and the bosonic network model with $\beta = 2.5$ (graph(d)). 
Empty (filled) symbols refer to $c_3(k)$ ($c_4(k)$).
Graph(a) shows data as obtained in November '97 (circles), January
'99 (squares) and the data taken in January '01 (triangles). Solid
lines refer to power law fittings, whose exponents are reported in
table (\ref{esponenti_delta}). 
}
\label{C3C4.fig}
\end{figure}

Similarly, $c_{4,i}$ measures the number of quadrilaterals passing through node $i$ 
divided by the number of possible quadrilaterals $Z_i$. 
This last quantity is the sum of all possible primary quadrilaterals $Z_i^p$ 
(where all vertices are nearest neighbors of node $i$) 
and all possible secondary quadrilaterals $Z_i^s$ 
(where one of the vertices is a second neighbor of node $i$).
If node $i$ has $k_i^{nn}$ second neighbors, 
$Z_i^p=k_i(k_i-1)(k_i-2)/2$ and $Z_i^s=k_i^{nn}k_i(k_i-1)/2$.
In Fig. \ref{C3C4.fig} (a) we plot $c_3(k)$, $c_4(k)$ 
for the Internet data at three different times (November 1997, 
January 1999 and January 2001) showing that the behavior of $c_3(k)$ 
and $c_4(k)$ is invariant with time and scales as 
\be
c_h(k)\sim k^{-\delta(h)}
\label{delta.eq}
\ee
with $\delta(3)=0.7(1)$ and $\delta(4)=1.1(1)$.

In Fig. $\ref{C3C4.fig}$, we compare the behavior of $c_3(k)$ and $c_4(k)$
in real Internet data and in the Internet models.
We found a similar behavior in the three networks model and in the Internet 
with the $c_3(k)$ and $c_4(k)$ of the Internet models scaling as 
$(\ref{delta.eq})$. Exponents, however, vary
significantly, as shown in Table \ref{esponenti_delta}. 

The fitness model reproduces the best the Internet clustering
scaling pattern.
Nevertheless, we observe that the number of triangles and
quadrilaterals in real data is much larger than in the fitness network.
Indeed, we have 
$c_3(10^3)\sim 10^{-2}$ and $c_4(10^3)\sim 10^{-4}$ in the AS network,
while in the fitness model $c_3(10^3)\sim 10^{-3}$ and $c_4(10^3)\sim
10^{-5}$.

\begin{table}[b] 
\begin{center}

\begin{tabular}[c]{|c|c|c|}
\hline
System & $\delta(3)$ & $\delta(4)$ \\
\hline
AS & $0.7 \pm 0.1$ & $1.1 \pm 0.01$ \\
Fitness & $0.67 \pm 0.01$ & $0.99 \pm 0.01$ \\
GNG (p=0.5) & $0.32 \pm 0.02$ & $1.68 \pm 0.03$ \\
GNG (p=0.6) & $0.27 \pm 0.02$ & $0.93 \pm 0.01$  \\
Bosonic ($\beta=2.5$) & $0.91 \pm 0.04$ & $1.07 \pm 0.07$ \\
\hline
\end{tabular}
\end{center}
\caption{The exponent of the clustering coefficient $c_3(k)$ and
$c_4(k)$ as measured from Internet data and from simulations of
network models.}
\label{esponenti_delta}
\end{table}


In conclusion, we computed the number $N_h(t)$ of $h$-loops of 
size $h=3,4,5$ in the Internet at the Autonomous System level and 
we have identified them as proper motifs of the Internet.
We have then compared the actual data with the behavior of $N_h(N)$ 
in the fitness model, in the GNG model and in the Bosonic network,
chosen as the most accurate Internet model developed to our best
knowledge.
Aside, the generalized clustering coefficients around individual nodes
have been investigated as a function of nodes degrees.
We have observed that, although some qualitative feature of the loop
scaling and of the clustering coefficient are captured by models, the
much larger number of cycles observed in the real network invoke for
improvement of the theory.

The authors are grateful to Uri Alon, Shalev Itzkovitz and Yi-Cheng Zhang  
for useful comments and discussions.
This paper has been financially supported by the Swiss National Foundation, under grant 
no. 2051-067733.02/1, and by the European Commission - FET Open project COSIN IST-2001-33555.

\end{document}